\pdfoutput=1
\documentclass[%
 reprint,
 amsmath,amssymb,
 aps,
 floatfix,
]{revtex4-2}

\usepackage{graphicx}
\usepackage{multirow}
\usepackage{bm}
\usepackage{algpseudocode}
\usepackage{hyperref}

\graphicspath{%
  {./}%
}

\newcommand{\TVThreshLambda}{9.4\times 10^{5}}
\newcommand{\TVThreshKappa}{0.75}

\begin{document}

\title{Verifying the High-Energy Tail in Bi-Kappa Particle Loading:\\
Three-Dimensional Tests, Velocity Bounds, and Finite Precision}

\author{Keheng Zhu}
 \email{zkeheng@umich.edu}
\affiliation{%
 University of Michigan, Ann Arbor, MI, USA%
}

\author{Yuri A. Omelchenko}
 \email{omelche@gmail.com}
\affiliation{%
 Space Science Institute, Boulder, CO, USA%
}

\date{\today}

\begin{abstract}
The bi-Kappa distribution models anisotropic space-plasma populations with
suprathermal power-law tails. Kinetic simulations represent such a population
by computational particles, and a loader draws their velocities from the
requested bi-Kappa distribution, which we call the target distribution.
When $\kappa$ is close to $1/2$, floating-point underflow can remove the
largest draws, and a velocity bound changes the loaded distribution. Published
Kappa loaders are usually verified by comparing histograms of the speed or of
one velocity component with the analytic distribution. Such comparisons do not
examine the direction of the velocity or whether the direction is independent
of the speed. We build an anisotropic loader on an established Gamma-ratio
construction and examine each of these problems. Near $\kappa=1/2$, evaluating
the Gamma ratio on a logarithmic scale avoids the loss of draws to underflow in
single and double precision down to $\kappa=0.5001$. When draws outside a bound
on each velocity component are discarded and redrawn, a small rejected fraction
can accompany a large change in the tail. At $\kappa=2$ and
$\theta_\parallel=2\theta_\perp$, a bound at $20$ times each component's
thermal speed rejects a fraction $3.6\times10^{-4}$ of the draws and lowers
the $99.9$th-percentile speed by $11\%$. We test the loaded velocities through
their direction and their normalized radius, the length of the velocity after
each component is divided by $\sqrt{\kappa}$ times its thermal speed. The
tests used nine values of $\kappa$ from $0.51$ to $10$, including
$1/2<\kappa\le3/2$, where the second moments do not exist, with
$\theta_\parallel=2\theta_\perp$ in double precision. At every $\kappa$,
$\chi^{2}$ tests find the normalized radius out to its $99.999$th percentile,
the direction, and their joint distribution consistent with the target
distribution.
The loader and a program that applies these tests to velocities from other
loaders are released as open-source software.
\end{abstract}

\maketitle

% Avoid stretched inter-paragraph glue on sparsely filled columns
% (otherwise REVTeX's flush-bottom justification opens visible gaps).
\raggedbottom

\section{Introduction}\label{sec:intro}

Collisionless and weakly collisional space plasmas often depart from Maxwellian
velocity distributions. Many contain suprathermal power-law tails that
influence particle energy, wave growth, instability thresholds, and
transport~\cite{livadiotis2013toolbox}. Kappa distributions are widely used to
model these tails~\cite{vasyliunas1968survey,pierrard2010kappa}. Their spectral index $\kappa$ controls the tail
weight. Smaller values give heavier tails, and the Maxwellian limit is recovered
as $\kappa\rightarrow\infty$.

Observed velocity distributions can also be anisotropic relative to the local
magnetic field. Solar-wind measurements show different parallel and
perpendicular widths together with suprathermal
tails~\cite{matteini2007evolution,hellinger2006anisotropy,pierrard2016electron}.
The bi-Kappa distribution considered here combines a Kappa tail with separate
parallel and perpendicular thermal speeds~\cite{summers1991modified,
mace1995dispersion,livadiotis2021anisotropic}. It is rotationally symmetric
about the magnetic-field direction.

Kinetic simulations approximate a bi-Kappa population with a finite set of
computational particles~\cite{birdsall1985plasma,hockney1981computer,
verboncoeur2005review}. We use \emph{loader} for the code that draws velocity
samples, applies the physical scales, enforces any requested velocity bound,
and rotates the samples into simulation coordinates. We call the bi-Kappa
distribution that a user requests the \emph{target distribution}. If the loaded
velocities do not follow the target distribution, the simulation starts from a
different distribution than the one intended. An incorrect load can raise the
field fluctuations above their expected level~\cite{cartwright2000loading}, and
a Kappa load with the wrong velocity-space shape can drive instabilities in a
plasma that should be stable~\cite{abdul2015bernstein}.

Drawing velocities from a Kappa distribution is harder than drawing them from
a Maxwellian, whose components are independent normal variates. The Kappa
cumulative distribution function has no elementary inverse, so the
inverse-transform method needs a numerical inverse or an approximation. The
algebraic tail is heavy, so a rejection sampler needs a proposal density with a
tail at least as heavy, and its acceptance rate depends on how closely this
proposal fits the target. Several established approaches solve these
problems. They include a polar transformation of two uniform variates for one
velocity component~\cite{abdul2014kappa}, numerical inverse
transforms~\cite{an2022chebsampling,zenitani2026approximate}, dedicated
rejection samplers~\cite{zenitani2025simple}, and scale
mixtures~\cite{abdul2015bernstein,zenitani2022relativistic,
zenitani2023losscone}. A scale mixture
combines a normal vector with an independent random scale derived from a Gamma
variate, so it needs only standard normal and Gamma generators. The
Gamma-ratio construction used here belongs to this class and is mathematically
equivalent to published constructions~\cite{abdul2015bernstein,
zenitani2022relativistic,zenitani2023losscone,zenitani2026nonmaxwellian}.

Published Kappa loaders are verified by comparing histograms of the speed, the
energy, or one velocity component with the analytic
distribution~\cite{abdul2014kappa,zenitani2022relativistic,zenitani2025simple,
zenitani2026approximate,zenitani2026nonmaxwellian}, in some cases together with
two-dimensional maps of the density~\cite{an2022chebsampling,
zenitani2026nonmaxwellian} or three-dimensional surfaces of constant
density~\cite{abdul2018thesis}. Recent work adds a binned Kullback--Leibler
divergence that decreases with sample size~\cite{zenitani2026approximate,
zenitani2026nonmaxwellian}. These comparisons are made mostly at $\kappa\ge2$.
To our knowledge, none of these studies tests the direction of the velocity or
whether the direction is independent of the speed, and none applies a
hypothesis test.

Building on the Gamma-ratio construction, we ask whether the complete
loader, including its anisotropic scaling, its rotation, and any velocity
bound, returns the target distribution in finite-precision arithmetic, and,
with a bound, the target distribution conditioned on the allowed region.
Three effects can change the loaded tail or escape the usual checks.

First, finite precision can alter the high-energy tail. The construction
draws the normalized radius $R$, the length of the velocity after each
component is divided by $\sqrt{\kappa}$ times its thermal speed, as a ratio of
two Gamma variates. The variate in the denominator has shape parameter
$\kappa-1/2$. As $\kappa$ approaches $1/2$, this variate can underflow to zero
in floating point~\cite{devroye1986nonuniform,liu2017gamma}, which makes $R$
infinite even when the final velocity components would fit in the
floating-point format. Zenitani and Nakano~\cite{zenitani2022relativistic},
whose parameterization by temperature restricts their method to $\kappa>3/2$
(Sec.~\ref{sec:conventions}), avoid a zero denominator by regenerating
the Gamma variate, by using a Gamma generator that never returns zero, or by
adding an offset of the order of the machine epsilon. Zenitani \emph{et
al.}~\cite{zenitani2026nonmaxwellian} note that for $\kappa\le3/2$ the shape
parameter $\kappa-1/2$ of the denominator variate is at most one, so that a
division by zero can occur, and recommend a piecewise rejection method in that
range. Near $\kappa=1/2$ the zeros come from the largest radii, so
discarding, redrawing, or offsetting these samples biases the high-energy
tail.

Second, a simulation may require bounded particle velocities, for example to
exclude superluminal speeds at low $\kappa$~\cite{scherer2019applicability} or
to respect a timestep limit~\cite{werner2018slpic}. The bound studied here,
which we call a \emph{cap}, is a hard limit on each velocity component in
field-aligned coordinates, whose axes are the magnetic-field direction and two
directions perpendicular to it. Any draw outside the cap is discarded and
redrawn. The retained draws then follow the \emph{capped distribution}, which
is the target distribution renormalized over the allowed region. The expected
rejected fraction equals the probability mass that the cap removes.
Section~\ref{sec:cap-tail} shows how much the high-speed tail can change when
this fraction is small.

Third, the three-dimensional distribution is not determined by the
distributions of the components or of the speed. Drawing the three components
independently from their one-dimensional distributions reproduces each
component but not the Kappa distribution~\cite{abdul2015bernstein}, and a
dependence between the speed and the direction can leave the speed
distribution unchanged. A test that detects such errors examines the direction
and its independence from the speed. At small $\kappa$ it also has to reach far
into the tail. For $1/2<\kappa\le3/2$ the second moments do not
exist~\cite{livadiotis2013toolbox,lazar2016interpretation}, so only statistics
that need no moments are available there.

After the background and related work of Sec.~\ref{sec:background}, the paper
has four parts, followed by the scope and limitations of the study
(Sec.~\ref{sec:discussion}) and a summary.

\begin{enumerate}
\item \emph{Probabilistic construction} (Sec.~\ref{sec:construction}): the
radius, the uniform direction, the anisotropic scaling, and the rotation into
simulation coordinates that together specify one uncapped draw.

\item \emph{Finite-precision behavior} (Sec.~\ref{sec:precision}): the loss
of draws near $\kappa=1/2$ in single and double precision, and a calculation
that keeps the radius on a logarithmic scale.

\item \emph{Finite velocity bounds and the loaded distribution}
(Sec.~\ref{sec:truncation}): the distribution of the retained draws, the
decay of the rejected fraction with the width of the bound, and the change of
a tail percentile.

\item \emph{Validation} (Sec.~\ref{sec:validation}): tests of the
distribution of $R$, of the direction and its independence from $R$, and of
invariance under changes of the thermal speeds and the field direction.
\end{enumerate}

\section{Background and related work}\label{sec:background}

\subsection{The Kappa and bi-Kappa distributions}\label{sec:kappa-dist}

We write probability density functions as $p(\cdot)$, with subscripts naming
the random variable. Random variables are capitals
($\mathbf{V},\mathbf{U},R,T$) and their realizations lowercase
($\mathbf{v},\mathbf{u},r,t$).

The isotropic Kappa distribution in three velocity dimensions is
\begin{equation}
f^{\mathrm{iso}}_{\kappa}(\mathbf{v})
=\frac{1}{\pi^{3/2}\theta^{3}}
\frac{\Gamma(\kappa+1)}{\kappa^{3/2}\,\Gamma\!\left(\kappa-\tfrac12\right)}
\left[1+\frac{|\mathbf{v}|^{2}}{\kappa\theta^{2}}\right]^{-(\kappa+1)},
\label{eq:kappa-iso}
\end{equation}
and its anisotropic generalization is the bi-Kappa distribution. We write it
in field-aligned coordinates, in which a velocity
$\mathbf{v}=(v_{\perp1},v_{\perp2},v_\parallel)$ has two Cartesian components
perpendicular to the magnetic-field direction $\hat{\mathbf{b}}$ and one along
it, with $v_\perp^{2}=v_{\perp1}^{2}+v_{\perp2}^{2}$:
\begin{widetext}
\begin{equation}
f_\kappa(\mathbf{v})
= \frac{1}{\pi^{3/2}\,\theta_\perp^{2}\,\theta_\parallel}\;
\frac{\Gamma(\kappa+1)}{\kappa^{3/2}\,\Gamma\!\left(\kappa-\tfrac{1}{2}\right)}
\left[\,1+\frac{v_\perp^{2}}{\kappa\,\theta_\perp^{2}}
      +\frac{v_\parallel^{2}}{\kappa\,\theta_\parallel^{2}}\,\right]^{-(\kappa+1)},
\qquad \kappa>\tfrac{1}{2}.
\label{eq:bi-kappa}
\end{equation}
\end{widetext}
Equation~\eqref{eq:kappa-iso} is the special case $\theta_\perp=\theta_\parallel=\theta$
of Eq.~\eqref{eq:bi-kappa}. Both are normalized to unity,
$\int f_\kappa\,\mathrm{d}^{3}v=1$, for every $\kappa>1/2$;
$\Gamma(\kappa-1/2)$ in the denominator requires this lower bound. Abdul and
Mace identified the isotropic Kappa distribution with a multivariate
Student-$t$ distribution~\cite{abdul2014kappa,abdul2015bernstein}. In the same
way, Eq.~\eqref{eq:bi-kappa} is a centered trivariate Student-$t$ distribution
with $\nu=2\kappa-1$ degrees of freedom and scale matrix
$\frac{\kappa}{2\kappa-1}\,\mathrm{diag}(\theta_\perp^{2},\theta_\perp^{2},\theta_\parallel^{2})$,
and the condition $\kappa>1/2$ is the condition $\nu>0$. As
$\kappa\rightarrow\infty$, Eq.~\eqref{eq:bi-kappa} converges to the
bi-Maxwellian
\begin{equation}
f_{M}(\mathbf{v})
= \frac{1}{\pi^{3/2}\,\theta_\perp^{2}\,\theta_\parallel}
\exp\!\left(-\frac{v_\perp^{2}}{\theta_\perp^{2}}
            -\frac{v_\parallel^{2}}{\theta_\parallel^{2}}\right).
\label{eq:bi-maxwell}
\end{equation}

Normalizability does not imply that all moments exist. The moment of order
$\ell$ converges only for
$\kappa>(\ell+1)/2$~\cite{livadiotis2013toolbox,lazar2016interpretation}, that
is, for $\ell<\nu$. The
second moments therefore exist only for $\kappa>3/2$, where
\begin{equation}
\mathrm{Var}(V_{\perp1})=\mathrm{Var}(V_{\perp2})=\frac{\theta_\perp^{2}\kappa}{2\kappa-3},
\qquad
\mathrm{Var}(V_\parallel)=\frac{\theta_\parallel^{2}\kappa}{2\kappa-3},
\label{eq:variances}
\end{equation}
diverging as $\kappa\rightarrow3/2^{+}$. For $1/2<\kappa\le3/2$, variance-based
verification is unavailable because there is no finite population value to
compare with a sample variance, and a population loaded in this range has no
finite kinetic temperature in the usual sense. In a nonrelativistic model the algebraic tail also extends to superluminal
speeds, and for $\kappa<2$ the superluminal part of the distribution can
contribute appreciably to the pressure~\cite{scherer2019applicability}, which
is one reason a loader may bound the velocities (Sec.~\ref{sec:truncation}).

\subsection{Parameters, conventions and units}\label{sec:conventions}

Table~\ref{table:notations} lists the main symbols with their physical meanings,
domain, and units. One convention needs stating explicitly because the Kappa
literature is not uniform. The parameters $\theta_\perp$ and $\theta_\parallel$ are \emph{thermal speeds}, not
temperatures, and have units of velocity. A common alternative convention writes
$\theta^{2}=2\left[(\kappa-3/2)/\kappa\right](k_{B}\mathcal{T}/m)$, which ties
$\theta$ to a kinetic temperature $\mathcal{T}$ but is defined only for
$\kappa>3/2$~\cite{abdul2014kappa,zenitani2022relativistic}. With thermal
speeds as parameters, the density of Eq.~\eqref{eq:bi-kappa} is normalizable
and the Gamma-ratio construction applies for all $\kappa>1/2$
(Sec.~\ref{sec:radial}), as in Ref.~\cite{zenitani2026nonmaxwellian}. Scherer
\emph{et al.}~\cite{scherer2020cookbook} compare the conventions and give the
moments of the Kappa variants in a common parameterization.

\begin{table}[t]
\caption{Symbols, physical meanings, domains, and units. $[v]$ denotes units of
velocity. The distribution is normalized to unit probability; a physical number
density $n$ enters as a multiplicative factor $n\,f_\kappa$ and does not
affect sampling.}
\label{table:notations}
\centering
\renewcommand{\arraystretch}{1.25}
\begin{ruledtabular}
\begin{tabular}{llll}
\textbf{Symbol} & \textbf{Meaning} & \textbf{Domain} & \textbf{Units} \\
\hline
$\kappa$ & spectral index & $(1/2,\infty)$ & --- \\
$\theta_\perp$ & perpendicular thermal speed & $(0,\infty)$ & $[v]$ \\
$\theta_\parallel$ & parallel thermal speed & $(0,\infty)$ & $[v]$ \\
$\hat{\mathbf{b}}$ & magnetic-field direction & $S^{2}$ & --- \\
$\mathbf{V}$ & velocity, field-aligned coordinates & $\mathbb{R}^{3}$ & $[v]$ \\
$v_{\perp1},v_{\perp2}$ & components perpendicular to $\hat{\mathbf{b}}$ & $\mathbb{R}$ & $[v]$ \\
$v_\parallel$ & component along $\hat{\mathbf{b}}$ & $\mathbb{R}$ & $[v]$ \\
$v_\perp$ & perpendicular speed & $[0,\infty)$ & $[v]$ \\
$\mathbf{V}_{\mathrm{sim}}$ & velocity, simulation coordinates & $\mathbb{R}^{3}$ & $[v]$ \\
$\mathbf{U}$ & normalized velocity, Eq.~\eqref{eq:u-def} & $\mathbb{R}^{3}$ & --- \\
$R=|\mathbf{U}|$ & normalized radius & $[0,\infty)$ & --- \\
$T=R^{2}$ & squared normalized radius & $[0,\infty)$ & --- \\
$\lambda$ & component-wise cap, Eq.~\eqref{eq:box} & $(0,\infty]$ & --- \\
$N$ & sample size & $\mathbb{N}$ & --- \\
\end{tabular}
\end{ruledtabular}
\end{table}

\subsection{Three related distributions}\label{sec:three-laws}

Three related distributions appear in this paper and in the literature it
draws on.

\begin{enumerate}
\item The \emph{bi-Kappa distribution}, Eq.~\eqref{eq:bi-kappa}, has unbounded
support. This is the target distribution throughout.

\item The \emph{capped bi-Kappa distribution}, Eq.~\eqref{eq:capped-law} below,
is Eq.~\eqref{eq:bi-kappa} conditioned on a component-wise box of half-width
$\lambda$ thermal speeds, and Sec.~\ref{sec:truncation} characterizes it. Its
bounded support makes all its moments finite.

\item The \emph{regularized Kappa distribution} of Scherer, Fichtner and
Lazar~\cite{scherer2017regularized} multiplies the Kappa density by a smooth
exponential cutoff, which makes all moments finite for $\kappa>0$ while keeping
the density continuous; its anisotropic form has a cutoff in each
direction~\cite{scherer2019anisotropic}. Unlike the hard cap of
Sec.~\ref{sec:truncation}, the cutoff is a parameter of the physical model,
alongside $\kappa$ and the thermal speed, and the standard Kappa distribution
is recovered as the cutoff parameter tends to zero~\cite{scherer2017regularized}.
\end{enumerate}

\subsection{Methods for generating Kappa variates}

Published approaches include the following three classes.

\emph{Inverse transform.} Sampling $F^{-1}(U)$ with $U\sim\mathcal{U}(0,1)$
requires the quantile function. For the radial variate used here the
cumulative distribution function is available in closed form as a regularized
incomplete beta function (Sec.~\ref{sec:radial}), but its inverse has no
elementary expression, so the method needs a numerical special-function
inverse or an approximation. An \emph{et al.}~\cite{an2022chebsampling}
make this practical in one and two dimensions with Chebyshev interpolation, and
Zenitani and Umeda~\cite{zenitani2026approximate} use a $q$-exponential
approximation to the Kappa cumulative distribution function.

\emph{Acceptance--rejection.} The method draws candidates from a proposal
density whose scaled version, the envelope, bounds the target, and accepts each
candidate with probability equal to the ratio of the target density to the
envelope~\cite{devroye1986nonuniform}. Tighter envelopes give higher
acceptance rates. In a comparison with their polar
transformation, Abdul and Mace~\cite{abdul2014kappa} measured acceptance
$\approx1/13$ with a rectangular envelope, while
Zenitani~\cite{zenitani2025simple} reaches $0.73$--$0.8$ with a Pareto envelope
on the same target, quoted for $\kappa\ge3/2$.
Zenitani \emph{et al.}~\cite{zenitani2026nonmaxwellian} note that
acceptance--rejection becomes less efficient in three dimensions when the
envelope is poorly matched.

\emph{Scale mixtures.} A Kappa variate can be written as a normal vector
divided by an independent Gamma-type variate, which is the multivariate
Student-$t$ representation.
Abdul and Mace~\cite{abdul2015bernstein} use a normal-over-$\chi^2$ mixture with
the $\chi^2$ deviate shared across components. Zenitani and
Nakano~\cite{zenitani2022relativistic} write it as a ratio of Gamma variates.
In a later paper~\cite{zenitani2023losscone}, they draw a Kappa loss-cone
distribution in the same way. Without the loss cone, this distribution is the
bi-Kappa distribution, with separate perpendicular and parallel thermal
speeds.
Zenitani \emph{et al.}~\cite{zenitani2026nonmaxwellian} generalize it to the
$(r,q)$ family, a two-parameter family of non-Maxwellian distributions that
contains the Kappa distribution as $r=0$, $q=\kappa+1$, and give an explicit
anisotropic loader. We use the Gamma-ratio scale mixture
for the uncapped loader. In exact arithmetic it returns the target
distribution for every $\kappa>1/2$ (Sec.~\ref{sec:construction}), with no
parameter that has to be chosen for each $\kappa$, and it is the generator used
in particle simulations by Refs.~\cite{abdul2015bernstein,
zenitani2022relativistic,zenitani2023losscone,zenitani2026nonmaxwellian}.
Each pass through the construction returns one draw, although the Gamma
generators it calls may use rejection internally; only the cap of
Sec.~\ref{sec:truncation} discards draws. The construction gives the radius as
an explicit ratio of two Gamma variates, and near $\kappa=1/2$ the variate in
the denominator can underflow to zero; Sec.~\ref{sec:precision} examines this
case. The capped distribution of Sec.~\ref{sec:truncation} and the tests of
Sec.~\ref{sec:validation} do not depend on how the uncapped draws are
generated.

\subsection{Relation to the isotropic problem}\label{sec:aniso-vs-iso}

Anisotropy does not change the radial sampling problem. As
Sec.~\ref{sec:isotropize} shows, an affine rescaling maps
Eq.~\eqref{eq:bi-kappa} onto the same dimensionless isotropic core as
Eq.~\eqref{eq:kappa-iso}. The radial distribution is therefore identical in the two cases
and depends only on $\kappa$. Anisotropy adds separate perpendicular and parallel scales, and a test has to
check each of them. It also adds a rotation from field-aligned into simulation
coordinates (Sec.~\ref{sec:transform}). A velocity-space bound is no longer
spherically symmetric, so its shape depends on whether it is defined in
field-aligned or in simulation coordinates (Sec.~\ref{sec:truncation}).

\section{Probabilistic construction}\label{sec:construction}

This section derives one draw from the uncapped target distribution,
Eq.~\eqref{eq:bi-kappa}: the anisotropic rescaling, the radial distribution,
the direction, and the rotation into simulation coordinates.
Section~\ref{sec:truncation} treats the capped distribution.

\subsection{Gamma and Beta-prime variates}\label{sec:betaprime}

We write $X\sim\Gamma(\alpha)$ for a unit-scale Gamma variate with shape
$\alpha>0$. If $X_1\sim\Gamma(a)$ and $X_2\sim\Gamma(b)$ are independent, their
ratio
\begin{equation}
T=\frac{X_1}{X_2}
\label{eq:t-x1x2-relation}
\end{equation}
follows the Beta-prime distribution $T\sim\beta'(a,b)$, with
density~\cite{devroye1986nonuniform}
\begin{equation}
p_T(t;a,b)=\frac{t^{a-1}(1+t)^{-(a+b)}}{\mathrm{B}(a,b)},\quad t>0,
\label{eq:betaprime-pdf}
\end{equation}
where $\mathrm{B}(a,b)=\Gamma(a)\Gamma(b)/\Gamma(a+b)$.

\subsection{Isotropization}\label{sec:isotropize}

In field-aligned coordinates, write the anisotropic rescaling in matrix form
as
$\mathbf{V}=A\mathbf{U}$ with
\begin{equation}
A=\mathrm{diag}\!\left(\sqrt{\kappa}\,\theta_\perp,\;
\sqrt{\kappa}\,\theta_\perp,\;\sqrt{\kappa}\,\theta_\parallel\right),
\qquad
\mathbf{U}=A^{-1}\mathbf{V},
\label{eq:u-def}
\end{equation}
so that componentwise
$U_{\perp1}=V_{\perp1}/(\sqrt{\kappa}\theta_\perp)$,
$U_{\perp2}=V_{\perp2}/(\sqrt{\kappa}\theta_\perp)$,
$U_\parallel=V_\parallel/(\sqrt{\kappa}\theta_\parallel)$.
Substituting $\mathbf{v}=A\mathbf{u}$ into Eq.~\eqref{eq:bi-kappa} reduces the
anisotropic quadratic form to an isotropic one:
\begin{equation}
1+\frac{v_{\perp1}^{2}+v_{\perp2}^{2}}{\kappa\theta_\perp^{2}}
+\frac{v_\parallel^{2}}{\kappa\theta_\parallel^{2}}
= 1+u_{\perp1}^{2}+u_{\perp2}^{2}+u_\parallel^{2}
= 1+|\mathbf{u}|^{2}.
\label{eq:quadratic-collapse}
\end{equation}
By the change-of-variables rule for densities, the linear map gives
$p_{\mathbf{U}}(\mathbf{u})=f_\kappa(A\mathbf{u})\,|\det A|$, with
\begin{equation}
|\det A| = \kappa^{3/2}\,\theta_\perp^{2}\,\theta_\parallel.
\label{eq:detA}
\end{equation}
The
factor $\kappa^{3/2}\theta_\perp^{2}\theta_\parallel$ from
Eq.~\eqref{eq:detA} cancels the corresponding factors in the prefactor of
Eq.~\eqref{eq:bi-kappa}, giving
\begin{equation}
p_{\mathbf{U}}(\mathbf{u})=\frac{\Gamma(\kappa+1)}{\pi^{3/2}\,\Gamma(\kappa-1/2)}
\left(1+|\mathbf{u}|^{2}\right)^{-(\kappa+1)},
\label{eq:pu}
\end{equation}
which depends on $\mathbf{u}$ only through $|\mathbf{u}|$ and is therefore
spherically symmetric. The parameters $\theta_\perp$ and $\theta_\parallel$ do
not appear in Eq.~\eqref{eq:pu}; their effect is included in the definition of
$\mathbf{U}$.

For $\theta_\perp\neq\theta_\parallel$, constant-density surfaces are
ellipsoids in physical velocity space, with semi-axes proportional to
$\sqrt{\kappa}\theta_\perp$, $\sqrt{\kappa}\theta_\perp$, and
$\sqrt{\kappa}\theta_\parallel$. Equation~\eqref{eq:u-def} divides each
coordinate by its scale and maps these ellipsoids to spheres in normalized
$\mathbf{u}$-space.

\subsection{The radial distribution}\label{sec:radial}

Because Eq.~\eqref{eq:pu} is spherically symmetric, the direction of
$\mathbf{U}$ is uniform on the unit sphere and independent of its length. Let
$R=|\mathbf{U}|$. Passing to spherical coordinates with polar angle $\Theta$
contributes the Jacobian $r^{2}\sin\Theta$, and integrating over the solid angle gives a factor
$4\pi r^{2}$, so the change-of-variables rule yields
\begin{equation}
\begin{aligned}
p_R(r)&=4\pi r^{2}\,p_{\mathbf{U}}(\mathbf{u})\big|_{|\mathbf{u}|=r}\\
&=\frac{4\,\Gamma(\kappa+1)}{\sqrt{\pi}\,\Gamma(\kappa-1/2)}\,
r^{2}\left(1+r^{2}\right)^{-(\kappa+1)}.
\end{aligned}
\label{eq:3D-radial-law}
\end{equation}
For $T=R^{2}$ the same rule applies with the scalar Jacobian
$\mathrm{d}r/\mathrm{d}t=1/(2\sqrt{t})$, so
\begin{equation}
\begin{aligned}
p_T(t)&=p_R(\sqrt{t})\,\frac{1}{2\sqrt{t}}
=\frac{2\,\Gamma(\kappa+1)}{\sqrt{\pi}\,\Gamma(\kappa-1/2)}\,
t^{1/2}(1+t)^{-(\kappa+1)}\\
&=\frac{t^{1/2}(1+t)^{-(\kappa+1)}}{\mathrm{B}(3/2,\;\kappa-1/2)},
\end{aligned}
\label{eq:pt}
\end{equation}
using $\Gamma(3/2)=\sqrt{\pi}/2$. Comparing Eq.~\eqref{eq:pt} with
Eq.~\eqref{eq:betaprime-pdf} identifies
\begin{equation}
T=R^{2}\sim\beta'\!\left(\tfrac32,\;\kappa-\tfrac12\right),
\label{eq:T-betaprime}
\end{equation}
so by Eq.~\eqref{eq:t-x1x2-relation} the radius can be generated from two
Gamma variates, $X_1\sim\Gamma(3/2)$ and $X_2\sim\Gamma(\kappa-1/2)$, as
$R=\sqrt{X_1}/\sqrt{X_2}$. The requirement $\kappa-1/2>0$ for the second shape
parameter is the same $\kappa>1/2$ that makes Eq.~\eqref{eq:bi-kappa}
normalizable.

Equations~\eqref{eq:T-betaprime} and~\eqref{eq:t-x1x2-relation} give the
construction used by Zenitani and Nakano~\cite{zenitani2022relativistic} and the
equivalent scale mixture of Abdul and Mace~\cite{abdul2015bernstein}. Without the
loss cone, the Kappa loss-cone loader of Ref.~\cite{zenitani2023losscone}
combines an equivalent scale mixture with the anisotropic rescaling of
Sec.~\ref{sec:isotropize}. The construction is the $r=0$, $q=\kappa+1$ case of the generalized construction
in Ref.~\cite{zenitani2026nonmaxwellian}, including its anisotropic rescaling.

The cumulative distribution function of $T$ is
\begin{equation}
F_T(t)=I_{t/(1+t)}\!\left(\tfrac32,\;\kappa-\tfrac12\right),
\label{eq:cdf}
\end{equation}
with $I$ the regularized incomplete beta function.

\subsection{The direction}\label{sec:direction}

A direction uniform on $S^{2}$ is obtained by drawing
$\cos\Theta\sim\mathcal{U}(-1,1)$ and $\Phi\sim\mathcal{U}(0,2\pi)$ and setting
\begin{equation}
\hat{\bm{\Omega}}=\left(\sin\Theta\cos\Phi,\;\sin\Theta\sin\Phi,\;\cos\Theta\right),
\qquad \mathbf{U}=R\,\hat{\bm{\Omega}},
\label{eq:direction}
\end{equation}
with components ordered as $(\perp1,\perp2,\parallel)$, so that $\Theta$ is the
angle between $\mathbf{U}$ and the field direction.
The solid-angle element is
$\mathrm{d}\Omega=\sin\Theta\,\mathrm{d}\Theta\,\mathrm{d}\Phi
=-\mathrm{d}(\cos\Theta)\,\mathrm{d}\Phi$. Thus $\cos\Theta$, rather than
$\Theta$, must be uniform. A uniform draw of $\Theta$ on $[0,\pi]$ places too
many points near the poles and does not produce the target distribution. The
same uniform distribution can be drawn without trigonometric functions by the
method of Marsaglia~\cite{marsaglia1972sphere}, which our implementation uses.

\subsection{Rotation into simulation coordinates}\label{sec:transform}

The steps above give the velocity $\mathbf{V}=A\mathbf{U}$, with components
$(V_{\perp1},V_{\perp2},V_\parallel)$ in field-aligned coordinates. A
simulation needs the velocity in its own Cartesian coordinates $x$, $y$, $z$,
in which the field direction $\hat{\mathbf{b}}=(b_x,b_y,b_z)$ generally varies
across the domain. We write the velocity in simulation coordinates as
$\mathbf{V}_{\mathrm{sim}}=(V_x,V_y,V_z)$.

The rotation requires two unit vectors $\hat{\mathbf{e}}_{\perp1}$ and
$\hat{\mathbf{e}}_{\perp2}$ that complete $\hat{\mathbf{b}}$ to a right-handed
orthonormal basis. Start with $\mathbf{t}=(1,1,1)$, let $k$ index the
largest-magnitude component of $\hat{\mathbf{b}}$, and replace $t_k$ by
$1-(b_x+b_y+b_z)/b_k$, so that $\mathbf{t}\cdot\hat{\mathbf{b}}=0$. Because
$|b_k|\ge1/\sqrt{3}$, this is well conditioned for every field direction. With
$\hat{\mathbf{e}}_{\perp2}=\mathbf{t}/\|\mathbf{t}\|$ and
$\hat{\mathbf{e}}_{\perp1}=\hat{\mathbf{e}}_{\perp2}\times\hat{\mathbf{b}}$,
\begin{equation}
\begin{aligned}
\mathbf{V}_{\mathrm{sim}}
&=V_{\perp1}\hat{\mathbf{e}}_{\perp1}+V_{\perp2}\hat{\mathbf{e}}_{\perp2}
+V_\parallel\hat{\mathbf{b}}
= Q\,\mathbf{V},\\
Q&=[\hat{\mathbf{e}}_{\perp1}\;\hat{\mathbf{e}}_{\perp2}\;\hat{\mathbf{b}}].
\end{aligned}
\label{eq:field-aligned-transform}
\end{equation}
For simplicity, our implementation uses $Q=I$ for the field direction
$(0,0,1)$, so that the two coordinate systems coincide.

Because $Q^{\mathsf{T}}Q=I$, the normalized velocity recovered from the
returned velocity, $A^{-1}Q^{\mathsf{T}}\mathbf{V}_{\mathrm{sim}}$, equals
$\mathbf{U}$, and its length equals the radius $R$ of Sec.~\ref{sec:radial}. Which perpendicular direction becomes
$\hat{\mathbf{e}}_{\perp1}$ does not affect the uncapped distribution, because
that distribution is rotationally symmetric about $\hat{\mathbf{b}}$. It does
set the orientation, in the plane perpendicular to the field, of the
component-wise velocity bound of Sec.~\ref{sec:truncation}. Because
$\hat{\mathbf{e}}_{\perp1}$ and $\hat{\mathbf{e}}_{\perp2}$ change abruptly
where the largest-magnitude component of $\hat{\mathbf{b}}$ changes, and
between $(0,0,1)$ and nearby directions, a bounded load has its allowed box
rotated about the field between neighboring field directions.

\subsection{Algorithm}\label{sec:algorithm}

Collecting Secs.~\ref{sec:isotropize}--\ref{sec:transform} gives the procedure
of Fig.~\ref{fig:algorithm}.

\begin{center}
\begin{minipage}{\columnwidth}
\hrule\vspace{2pt}\hrule\vspace{4pt}
\begin{algorithmic}[1]
\State $X_1\sim\Gamma(3/2)$, \quad $X_2\sim\Gamma(\kappa-1/2)$
\State $R\leftarrow\sqrt{X_1}/\sqrt{X_2}$
\State $\hat{\bm{\Omega}}\sim$ uniform on the unit sphere
\State \textbf{return} $\mathbf{V}_{\mathrm{sim}}=R\,Q\,A\,\hat{\bm{\Omega}}$
\end{algorithmic}
\vspace{4pt}\hrule
\refstepcounter{figure}\label{fig:algorithm}
\smallskip
{\small FIG.~\thefigure. One uncapped bi-Kappa draw for $\kappa>1/2$ and
$\theta_\perp,\theta_\parallel>0$. Lines 1--2 draw the radius
(Sec.~\ref{sec:radial}) and line 3 the direction (Sec.~\ref{sec:direction}).
Line 4 applies the anisotropic scaling $A$ of Eq.~\eqref{eq:u-def} and the
rotation $Q$ of Eq.~\eqref{eq:field-aligned-transform}. Our implementation
evaluates lines 1--2 on a logarithmic scale
(Sec.~\ref{sec:precision}). With a finite bound $\lambda$
(Sec.~\ref{sec:truncation}), a draw is kept only if
$\sqrt{\kappa}\,R\max_j|\hat{\Omega}_j|\le\lambda$; otherwise the procedure
returns to line 1.}
\end{minipage}
\end{center}

In exact arithmetic, Fig.~\ref{fig:algorithm} returns a velocity whose
field-aligned components follow the uncapped density of
Eq.~\eqref{eq:bi-kappa}. We call one pass through lines 1--4 an
\emph{attempt}. Without a cap, each attempt returns one draw. With a cap, an
attempt whose velocity lies outside the allowed region is discarded, so the
number of attempts can exceed the number of returned draws.
Section~\ref{sec:precision} measures the effects of floating-point
arithmetic.

\section{Finite-precision behavior}\label{sec:precision}

The radial sampler divides by the Gamma variate
$X_2\sim\Gamma(\kappa-1/2)$. As $\kappa\rightarrow1/2^{+}$, the shape
parameter approaches zero, and direct floating-point generation of $X_2$ can
underflow before the final velocity is formed. The calculation then fails even
when every final velocity component would fit in the output
floating-point format. Low $\kappa$ also gives the target distribution a
heavier high-energy tail, so some final velocity components are too large to
store at all. Both outcomes occur at large radii, so discarding or redrawing
the failed attempts depletes the high-energy tail, which matters in many space
plasma applications~\cite{pierrard2010kappa} and can alter wave growth in
kinetic plasma models~\cite{thorne1991enhancement}. Unlike an intentional
bound (Sec.~\ref{sec:truncation}), such a failure is not specified in advance,
and because it depends on the radius, the retained draws no longer follow the
target distribution. We separate the two outcomes and test whether a
calculation that keeps the radius on a logarithmic scale removes the first,
which is avoidable.

Generating Gamma variates with a small shape parameter is numerically
difficult. The standard way to extend a Gamma generator to shapes below one
uses the distributional identity $X_2=G\xi^{1/\alpha}$, which expresses a
Gamma variate of small shape $\alpha=\kappa-1/2$ through one of shape
$\alpha+1$, with $G\sim\Gamma(\alpha+1)$ and
$\xi\sim\mathcal{U}(0,1)$~\cite{devroye1986nonuniform}. Marsaglia and Tsang
extend their generator to small shapes in this
way~\cite{marsaglia2000simple}, and the default Gamma generator of MATLAB is
based on theirs~\cite{liu2017gamma}. Generating the variate on a logarithmic
scale avoids its underflow~\cite{liu2017gamma}. We compare two calculations of
the radius that use this identity. The \emph{direct calculation} forms $X_2=G\xi^{1/\alpha}$ as a floating-point product and then
evaluates
\begin{equation}
R=\frac{\sqrt{X_1}}{\sqrt{X_2}}.
\label{eq:direct-radius}
\end{equation}
For small $\alpha$ the product can round to zero, and $R$ is then non-finite
before the final velocity components are formed.

The \emph{stabilized calculation} keeps the logarithmic scale from the Gamma
variate to the final velocity components. It never forms $X_2$. From the same
$G$ and $\xi$ it evaluates $\log X_2=\log G+(\log\xi)/\alpha$ and then
\begin{equation}
\log R=\frac{\log X_1-\log X_2}{2}.
\label{eq:log-radius}
\end{equation}
It forms the final velocity components $V_j=R\,g_j$ from $\log R$, where $\mathbf{g}=QA\hat{\bm{\Omega}}$ is the scaled
and rotated direction of Fig.~\ref{fig:algorithm}. When $R$ is representable,
each component is $V_j=e^{\log R}g_j$. If $R$ itself exceeds the
largest representable number, each component is instead evaluated as
$V_j=\mathrm{sgn}(g_j)\exp(\log R+\log|g_j|)$. In exact arithmetic this
gives the same radius as Eq.~\eqref{eq:direct-radius}.

\begin{figure*}[t]
\centering
\includegraphics[width=\textwidth]{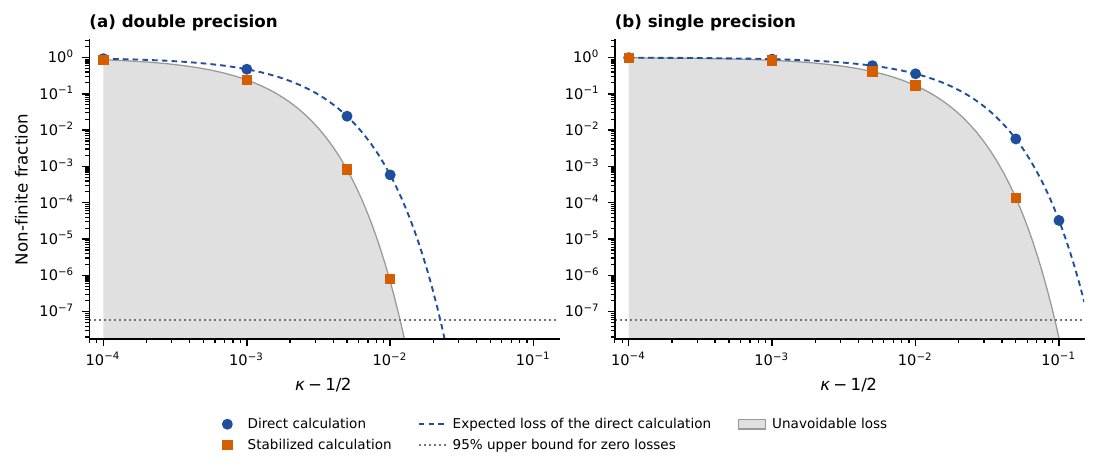}
\caption{Fraction of attempts that returned a non-finite velocity in (a)
double and (b) single precision, plotted against $\kappa-1/2$, for
$\theta_\perp=\theta_\parallel=1$ and the field along $\hat{\mathbf{z}}$. Each
marker is measured from $5\times10^7$ attempts without a velocity bound.
Settings with no non-finite output cannot be plotted on the logarithmic axis;
for them the fraction lies below the dotted line, the one-sided 95\% upper
bound for zero events in $5\times10^7$ attempts. The grey region is the fraction of
draws with a velocity component larger than the largest number the
floating-point format can store, which no calculation can return. The dashed
line is the probability that the product $X_2=G\xi^{1/\alpha}$ of the direct
calculation rounds to zero, and the direct calculation loses exactly these
draws. The losses of the stabilized calculation follow the upper edge of the
grey region. The gap between the dashed line and the grey region is the
avoidable loss, the draws whose velocity fits the format but whose product
rounds to zero.}
\label{fig:fp-envelope}
\end{figure*}

We tested both calculations in double and single precision at thirteen values,
$\kappa\in\{0.5001,\allowbreak 0.501,\allowbreak 0.505,\allowbreak 0.51,\allowbreak 0.55,\allowbreak 0.6,\allowbreak 0.75,\allowbreak 1,\allowbreak 1.25,\allowbreak 1.49,\allowbreak 1.5,\allowbreak 2,\allowbreak 5\}$,
with $\theta_\perp=\theta_\parallel=1$, the field along $\hat{\mathbf{z}}$, and
no velocity bound. The values $1.49$ and $1.5$ lie on either side of
$\kappa=3/2$, where the shape $\kappa-1/2$ of the denominator variate crosses
one, the lower limit of generators such as that of Marsaglia and
Tsang~\cite{marsaglia2000simple}. In
each attempt the two calculations received the same random variates, so their
outcomes can be compared attempt by attempt. Each setting used $5\times10^7$
attempts. Non-finite outputs occurred only for $\kappa\le0.6$, so
Fig.~\ref{fig:fp-envelope} shows this range.

Figure~\ref{fig:fp-envelope} compares the measured losses with two
probabilities, each computed in closed form. The grey region is the
probability that the exact velocity has a component larger than the largest
number of the floating-point format. It follows from the radial distribution,
Eq.~\eqref{eq:3D-radial-law}, and the uniform direction, and we call it the
\emph{unavoidable loss}, because no calculation that returns values in this
format can return these draws. The dashed curve is the probability that the
floating-point product $G\xi^{1/\alpha}$ of the direct calculation rounds to
zero.

The dashed curve is the whole loss of the direct calculation. A zero product
makes $R$ infinite, while a nonzero product keeps $R$ below about $10^{162}$
and every component finite. The direct calculation therefore loses a draw
exactly when the product rounds to zero. Its measured losses differ from the
dashed curve by less than $0.7\%$ of its value at every $\kappa$ with a
nonzero loss. The draws of the unavoidable loss are among them. Because $X_1$
is of order one, Eq.~\eqref{eq:direct-radius} gives $R\approx X_2^{-1/2}$. A
draw with a component too large for double precision has
$R>1.8\times10^{308}$, so its $X_2$ is about $10^{-616}$, far below the
smallest positive number, $4.9\times10^{-324}$. The dashed curve therefore
lies above the grey region.

In both precisions, the measured losses of the stabilized calculation lie on
the upper edge of the grey region at every $\kappa$ with a nonzero loss. The
stabilized calculation therefore loses only the unavoidable draws. The gap
between the dashed curve and the grey region is the avoidable loss. It
consists of draws whose exact velocity fits the format but whose product
rounds to zero, with radii between about $10^{162}$ and $1.8\times10^{308}$ in
double precision. In double precision at $\kappa=0.51$, for example, the
direct calculation lost $5.9\times10^{-4}$ of attempts and the stabilized
calculation $8.0\times10^{-7}$. Both losses are larger in single precision, because its
range is narrower. Its largest number is $3.4\times10^{38}$ and its smallest
positive number $1.4\times10^{-45}$, against $1.8\times10^{308}$ and
$4.9\times10^{-324}$ in double precision.

Close to $\kappa=1/2$ the loss that remains is set by the format. At
$\kappa=0.5001$, $87\%$ of attempts in double precision and $98\%$ in single
precision have a component too large for the format, so few draws can be
returned there by any calculation. Any practical cap lies far below the largest
number of the format. A capped loader therefore discards and redraws such a
draw like any other draw outside the allowed region, and these losses do not
change the capped distribution (Sec.~\ref{sec:truncation}).
Section~\ref{sec:validation} tests the distribution of the stabilized output
in double precision for $\kappa\ge0.51$.

\section{Finite velocity bounds and the loaded distribution}\label{sec:truncation}

This section treats the cap introduced in Sec.~\ref{sec:intro}. Cartwright,
Verboncoeur and Birdsall~\cite{cartwright2000loading} quantified the numerical
error of particle loads that include cutoff Maxwellians. A grid-based kinetic
code truncates the distribution at the edge of its velocity domain. Its
projected moments can match those of the target distribution while the
distribution held on the velocity grid does not~\cite{johnson2025moment}. We identify the distribution of the retained
draws, measure its distance from the target distribution, and quantify the
change in the high-speed tail.

\subsection{The capped distribution}\label{sec:capped}

With a finite cap $\lambda$, draws whose components, in units of their
thermal speeds, fall outside the box
\begin{equation}
B_\lambda=\left\{\mathbf{v}:\;
\frac{|v_{\perp1}|}{\theta_\perp}\le\lambda,\;
\frac{|v_{\perp2}|}{\theta_\perp}\le\lambda,\;
\frac{|v_\parallel|}{\theta_\parallel}\le\lambda\right\}
\label{eq:box}
\end{equation}
are discarded and redrawn. The retained draws are distributed as $\mathbf{V}$
conditioned on $\mathbf{V}\in B_\lambda$, that is, as the bi-Kappa distribution
truncated to $B_\lambda$:
\begin{equation}
\begin{aligned}
f_\lambda(\mathbf{v})
&=\frac{f_\kappa(\mathbf{v})\,\mathbf{1}\{\mathbf{v}\in B_\lambda\}}
{P(\mathbf{V}\in B_\lambda)}
=\frac{f_\kappa(\mathbf{v})}{P_\lambda}\quad\text{on }B_\lambda,\\
P_\lambda&\equiv P(\mathbf{V}\in B_\lambda).
\end{aligned}
\label{eq:capped-law}
\end{equation}
The allowed velocity-space region therefore determines the loaded
distribution together with $\kappa$, $\theta_\perp$, and $\theta_\parallel$.
Every moment of $f_\lambda$ is finite because its support is bounded. Because
$f_\kappa$ is a multivariate Student-$t$ distribution
(Sec.~\ref{sec:kappa-dist}), $f_\lambda$ is a multivariate $t$ distribution
truncated to a rectangle, whose
moments have closed forms and software
implementations~\cite{ho2012truncated,lin2024moments,galarza2021momtrunc}.
Moments computed from a capped load are moments of $f_\lambda$; for
$\kappa\le3/2$ the second moments grow without bound as $\lambda$ increases.

The bound acts on the field-aligned components, before the rotation of
Sec.~\ref{sec:transform}. In the normalized coordinates of
Eq.~\eqref{eq:u-def} it reduces to $\sqrt{\kappa}\max_j|u_j|\le\lambda$, with
$j$ running over $\perp1$, $\perp2$, and $\parallel$. The thermal speeds cancel,
so $P_\lambda$ and the conditional distribution of $\mathbf{U}$ depend on
$(\kappa,\lambda)$ alone. The capped distribution of $\mathbf{V}$ retains its dependence on
$\theta_\perp$ and $\theta_\parallel$ through the inverse scaling. With
$\mathbf{U}=R\hat{\bm{\Omega}}$ and $M=\max_j|\hat{\Omega}_j|$, a draw is
retained when $R^{2}\le\lambda^{2}/(\kappa M^{2})$. Because the direction is
uniform and independent of $R$, we compute $P_\lambda$ by averaging
Eq.~\eqref{eq:cdf} at this threshold over the distribution of $M$, a
one-dimensional numerical quadrature. At $\kappa=2$ and $\lambda=3$ the
quadrature gives $1-P_\lambda=0.0808$, and in $5\times10^{5}$ attempts the
capped sampler rejected a fraction $0.0812$, with a standard error of
$0.0004$.

\subsection{Rejected fraction and the high-speed tail}\label{sec:cap-tail}

The \emph{rejected fraction} $1-P_\lambda$, the probability that a draw is
rejected, equals the total-variation distance between the target distribution
and the capped distribution,
\begin{equation}
\mathrm{TV}(f_\lambda,f_\kappa)
=\tfrac12\int\left|f_\lambda-f_\kappa\right|\mathrm{d}^{3}v
= 1-P_\lambda.
\label{eq:tv}
\end{equation}
For every region of velocity space, the probabilities of that region under
$f_\lambda$ and under $f_\kappa$ therefore differ by at most $1-P_\lambda$. The
distance does not, by itself, bound the shift of a tail quantile in velocity
units.

\begin{figure*}
    \centering
    \includegraphics[width=1\linewidth]{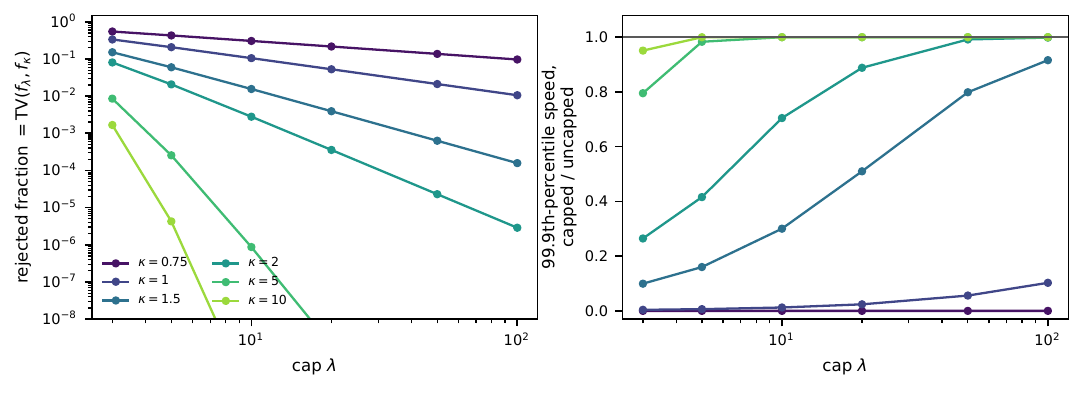}
    \caption{Effect of the component-wise cap on the rejected fraction and the
    high-speed tail. \emph{Left:} The rejected fraction $1-P_\lambda$, the
    probability of the target distribution outside the cap, computed by
    the quadrature of Sec.~\ref{sec:capped}. \emph{Right:} The capped-to-uncapped ratio of
    the $99.9$th-percentile speed; values closer to unity indicate better
    preservation of this percentile. The right-panel points are means over
    five pairs of capped and uncapped runs, each pair using the same random
    numbers, with $10^{5}$ accepted draws per run. Both panels use
    $(\theta_\perp,\theta_\parallel)=(1,2)$, six $\kappa$ values, six $\lambda$
    values, and double precision.}
    \label{fig:cap}
\end{figure*}

For the algebraic bi-Kappa tail considered here, the radial distribution of
the target, Eq.~\eqref{eq:3D-radial-law}, has
$P(R>r)\sim r^{-(2\kappa-1)}$. The rejected fraction therefore decays only
algebraically,
\begin{equation}
1-P_\lambda \;\sim\; \lambda^{-(2\kappa-1)},
\label{eq:tv-scaling}
\end{equation}
with the slowest decay for the heaviest tails. At $\kappa=0.75$, a cap at
$\lambda=100$ has a rejected fraction of $9.7\%$. Because the uncapped second
moment does not exist for $1/2<\kappa\le3/2$, we use the
$99.9$th-percentile speed as a common tail diagnostic across the full $\kappa$
range. The left panel of
Fig.~\ref{fig:cap} shows the rejected fraction across $\kappa$ and $\lambda$,
while the right panel shows how the cap changes the $99.9$th-percentile speed.

At $\kappa=2$ and $(\theta_\perp,\theta_\parallel)=(1,2)$, a cap at
$\lambda=20$ has a rejected fraction of $3.6\times10^{-4}$ and reduces the
$99.9$th-percentile speed by $11\%$. A small rejected fraction therefore does
not by itself guarantee that a tail statistic is preserved. For
$\kappa\le3/2$, all caps up to $\lambda=100$ in Table~\ref{table:cap} have a
rejected fraction above $10^{-4}$ and lower the $99.9$th-percentile speed by
more than $8\%$. At $\kappa=\TVThreshKappa$, the quadrature of
Sec.~\ref{sec:capped} gives a rejected fraction below $10^{-3}$ only for
$\lambda\gtrsim\TVThreshLambda$. Table~\ref{table:cap} lists both quantities
for each $(\kappa,\lambda)$.

The effect of a cap depends on the quantity of interest. A bound
well above the few thermal speeds that control a resonant instability may have
little effect on its linear growth
rate~\cite{trotignon1977cutoff,stucchi2025landau}, whereas a
quantity weighted toward the tail depends on the removed particles.

\begin{table*}[t]
\caption{Rejected fraction of the cap and preservation of the
$99.9$th-percentile speed. Each entry has the form rejected fraction / ratio.
The first number is the rejected fraction, computed by the quadrature of
Sec.~\ref{sec:capped}; the second is the capped-to-uncapped ratio of the
$99.9$th-percentile speed. A rejected fraction near zero means that the cap
removes few draws, and a ratio near unity means that the cap leaves the
$99.9$th-percentile speed nearly unchanged. Percentile ratios are averaged over five pairs of
capped and uncapped runs, each pair using the same random numbers, with
$10^{5}$ accepted draws per run for $(\theta_\perp,\theta_\parallel)=(1,2)$ in
double precision.}
\label{table:cap}
\centering
\setlength{\tabcolsep}{3pt}
\begin{ruledtabular}
\begin{tabular}{lcccccc}
$\kappa$ & $\lambda=3$ & $\lambda=5$ & $\lambda=10$ & $\lambda=20$ & $\lambda=50$ & $\lambda=100$ \\
\hline
$0.75$ & $0.55$ / $<\!0.001$ & $0.43$ / $<\!0.001$ & $0.31$ / $<\!0.001$ & $0.22$ / $<\!0.001$ & $0.14$ / $<\!0.001$ & $0.097$ / $<\!0.001$ \\
$1$ & $0.34$ / $0.004$ & $0.21$ / $0.006$ & $0.11$ / $0.012$ & $0.053$ / $0.024$ & $0.021$ / $0.056$ & $0.011$ / $0.103$ \\
$1.5$ & $0.15$ / $0.099$ & $0.060$ / $0.160$ & $0.016$ / $0.300$ & $0.0039$ / $0.510$ & $6.3\times 10^{-4}$ / $0.799$ & $1.6\times 10^{-4}$ / $0.916$ \\
$2$ & $0.081$ / $0.265$ & $0.021$ / $0.416$ & $0.0028$ / $0.705$ & $3.6\times 10^{-4}$ / $0.888$ & $2.3\times 10^{-5}$ / $0.992$ & $2.9\times 10^{-6}$ / $0.999$ \\
$5$ & $0.0086$ / $0.795$ & $2.6\times 10^{-4}$ / $0.983$ & $8.7\times 10^{-7}$ / $1.000$ & $2.0\times 10^{-9}$ / $1.000$ & $5.4\times 10^{-13}$ / $1.000$ & $1.1\times 10^{-15}$ / $1.000$ \\
$10$ & $0.0017$ / $0.951$ & $4.3\times 10^{-6}$ / $1.000$ & $7.2\times 10^{-11}$ / $1.000$ & $2.6\times 10^{-16}$ / $1.000$ & $8.7\times 10^{-24}$ / $1.000$ & $1.7\times 10^{-29}$ / $1.000$ \\
\end{tabular}
\end{ruledtabular}
\end{table*}

These results apply to the component-wise box of Eq.~\eqref{eq:box}. Because
the box is defined in units of each component's thermal speed, its rejected
fraction depends on $\kappa$ and $\lambda$ alone. A bound on the speed
$|\mathbf{v}|$ does not have this property. When
$\theta_\perp\neq\theta_\parallel$, its rejected fraction depends on both
thermal speeds, and truncation of this kind changes the ratio of the second
moments~\cite{tallis1963elliptical,palombi2017spherical}.

\section{Validation}\label{sec:validation}

\subsection{Design}

For each returned draw we compute the normalized velocity
$\mathbf{u}=A^{-1}Q^{\mathsf{T}}\mathbf{V}_{\mathrm{sim}}$, which undoes the
rotation and the anisotropic scaling of Secs.~\ref{sec:isotropize}
and~\ref{sec:transform}. Under the target distribution, $R=|\mathbf{u}|$ has
$R^{2}$ distributed as in Eq.~\eqref{eq:T-betaprime}, the direction of
$\mathbf{u}$ is uniform on the sphere, and the direction is independent of $R$
(Sec.~\ref{sec:radial}). These three properties determine the distribution, so
we test them. Beran~\cite{beran1979ellipsoidal} tests ellipsoidal
symmetry by the same reduction to a transformed radius and a direction, and
later work develops tests of spherical symmetry and of radial--directional
independence~\cite{henze2014spherical,zhang2026radial}. Here the scaling and the
radial distribution are known, so the test reduces to counting draws in cells
of equal probability (Sec.~\ref{sec:cell-test}). Because $\mathbf{u}$ is computed with the specified
thermal speeds and field direction, an error in the loader's scaling or
rotation would also appear as a violation of these properties. None of them
requires a finite moment, so the tests apply for $1/2<\kappa\le3/2$.

In runs with identical random numbers and caps from $\lambda=3$ to $100$, our
capped sampler returned exactly the uncapped draws that lie inside
$B_\lambda$. A capped sample therefore follows Eq.~\eqref{eq:capped-law}
whenever the uncapped sample follows the target, and we test our uncapped
sampler, in double precision. We drew samples at nine values,
$\kappa\in\{0.51,0.55,0.75,1,1.25,1.5,2,5,10\}$, with
$(\theta_\perp,\theta_\parallel)=(1,2)$ and the field along $\hat{\mathbf{z}}$,
for which the returned components are the field-aligned components
$(V_{\perp1},V_{\perp2},V_\parallel)$ of Sec.~\ref{sec:transform}. The general
rotation of Eq.~\eqref{eq:field-aligned-transform} is tested in
Sec.~\ref{sec:rotation-test}. Each
$\kappa$ was sampled with $5\times10^{7}$ draws from the Mersenne Twister
random-number generator~\cite{matsumoto1998mersenne}, and every test in this
section uses these draws. At $\kappa=0.51$, $39$ of the $5\times10^{7}$ draws
had infinite components; at the other values of $\kappa$ no draw had a
non-finite component. Under the target
distribution, a component exceeds the largest double-precision number with
probability $6.9\times10^{-7}$ at this $\kappa$, the unavoidable loss of
Sec.~\ref{sec:precision} for these thermal speeds, so $34$ such draws are
expected. The test of the radius in Sec.~\ref{sec:cell-test} counts these
draws in its outermost shell, and the test over all cells omits them, because
their direction is undefined.

\subsection{Radius, direction, and their independence}\label{sec:cell-test}

We divide the space of $\mathbf{u}$ into cells that the target distribution
fills with equal probability, the choice analyzed by Mann and
Wald~\cite{mann1942choice}, and compare the number of draws in each cell with
its expected value. Nine radii $r_1<\dots<r_9$, obtained from
Eq.~\eqref{eq:cdf}, divide $R$ into ten shells of probability $1/10$ each. The
direction is divided into $40$ cells by five equal intervals of $\cos\Theta$
and eight equal intervals of $\Phi$, where $\Theta$ and $\Phi$ are the polar
angle and the azimuth of $\mathbf{u}$ about the field. These cells have equal
solid angle because $\mathrm{d}\Omega=-\mathrm{d}(\cos\Theta)\,\mathrm{d}\Phi$
(Sec.~\ref{sec:direction}). The target makes the direction uniform and
independent of $R$, so each of the $10\times40=400$ combined cells has
probability $1/400$.

The outermost shell, $R>r_9$, holds the fastest $10\%$ of draws, since $r_9$
is the $90$th percentile of $R$. These draws determine tail statistics such as
the $99.9$th-percentile speed of Sec.~\ref{sec:cap-tail}. For the test of the
radius alone, we therefore divide this shell further at the $99$th, $99.9$th,
$99.99$th, and $99.999$th percentiles of $R$, which gives $14$ shells.

The shell edges span a wide range of radii. The $99.999$th percentile of $R$ is
$55$ at $\kappa=2$ and $1.4\times10^{250}$ at $\kappa=0.51$, and at
$\kappa=0.51$ about one draw in $1200$ has $R>1.3\times10^{154}$, where $R^{2}$
exceeds the largest double-precision number. Each draw is therefore assigned to
its shell by comparing $\log R$ with the logarithms of the shell edges.

The comparison uses Pearson's $\chi^{2}$ statistic~\cite{pearson1900chi2},
the sum over cells of $(O-E)^{2}/E$, where $O$ is the observed number of draws
in a cell and $E$ the expected number. For a sample from the target
distribution with large $E$,
this statistic follows the $\chi^{2}$ distribution with $m-1$ degrees of
freedom, where $m$ is the number of cells. The $p$-value is the probability
that this distribution exceeds the observed value of the statistic. A small $p$-value indicates a discrepancy, and we treat a $p$-value below
$0.01$ as one. A radial
distribution of the wrong shape changes the counts in the inner or the outer
shells. A direction that is not uniform, such as the clustering near the poles
described in Sec.~\ref{sec:direction}, changes the counts in particular
direction cells in every shell. A dependence between radius and direction
changes the counts in particular direction cells of particular shells. An
error in the thermal speeds or the rotation distorts $\mathbf{u}$ and changes
the counts in the same ways. The test over all $400$ cells has $399$ degrees of
freedom. The test of the radius alone compares the counts in the $14$ shells,
whose probabilities differ; it has $13$ degrees of freedom, and its smallest
expected count, beyond the $99.999$th percentile, is $500$.

\begin{table}[tb]
\caption{Comparison of the loaded velocities with the target distribution.
Each entry is the $p$-value of Pearson's $\chi^{2}$ test
(Sec.~\ref{sec:cell-test}); a value below $0.01$ would indicate a
discrepancy. The first column tests the radius alone, over the $14$ shells of
Fig.~\ref{fig:shells}. The second tests the radius and the direction together,
over $400$ cells of equal probability: ten radius shells of probability $1/10$
by $40$ direction cells. Rows above the line have $\kappa\le3/2$, where the
second moments do not exist. No velocity bound,
$(\theta_\perp,\theta_\parallel)=(1,2)$, field along $\hat{\mathbf{z}}$,
double precision, $5\times10^{7}$ draws at each $\kappa$.}
\label{table:validation}
\centering
\begin{ruledtabular}
\begin{tabular}{lcc}
\multirow{2}{*}{$\kappa$} & \multicolumn{2}{c}{$p$-value} \\
\cline{2-3}
 & Radius (14 shells) & Radius and direction (400 cells) \\
\hline
$0.51$ & 0.995 & 0.353 \\
$0.55$ & 0.073 & 0.062 \\
$0.75$ & 0.153 & 0.125 \\
$1$ & 0.671 & 0.464 \\
$1.25$ & 0.178 & 0.968 \\
$1.5$ & 0.532 & 0.992 \\
\hline
$2$ & 0.737 & 0.147 \\
$5$ & 0.356 & 0.672 \\
$10$ & 0.622 & 0.719 \\
\end{tabular}
\end{ruledtabular}
\end{table}

\begin{figure}[tb]
    \centering
    \includegraphics[width=\columnwidth]{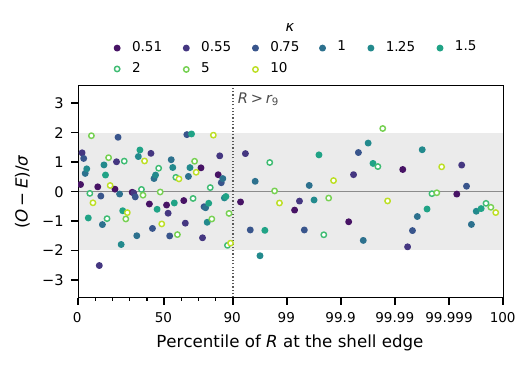}
    \caption{Number of loaded draws in each radius shell compared with the
    number expected under the target distribution. The radius is
    $R=|\mathbf{u}|$, the length of the normalized velocity, obtained by undoing
    the rotation and the anisotropic scaling (Sec.~\ref{sec:validation}). It is divided
    into $14$ shells: nine of probability $1/10$ inside $r_9$, the $90$th
    percentile of $R$, and five beyond it (right of the dotted line), bounded by
    the $99$th, $99.9$th, $99.99$th, and $99.999$th percentiles. The horizontal
    axis gives the percentile of $R$ under the target distribution at each shell
    edge; it is not to scale, and the five outer shells are drawn wider. For
    each shell, $O$ is the observed count, $E$ the expected count (the number of
    draws, $N=5\times10^{7}$, times the probability of the shell), and
    $\sigma=\sqrt{E(1-E/N)}$ the standard deviation of $O$ for a sample from
    the target distribution. For such a sample each value of $(O-E)/\sigma$ is
    approximately a
    standard normal variable, so about $95\%$ of the values should lie in the
    shaded band between $-2$ and $2$. No $\kappa$ departs systematically in any
    range of $R$. The
    markers of the nine $\kappa$ are placed side by side within each shell;
    filled markers have $\kappa\le3/2$, where the second moments do not exist.
    No velocity bound, $(\theta_\perp,\theta_\parallel)=(1,2)$, field along
    $\hat{\mathbf{z}}$, double precision.}
    \label{fig:shells}
\end{figure}

Table~\ref{table:validation} gives the $p$-values at each $\kappa$, with
$E=1.25\times10^{5}$ in each of the $400$ cells. None
is below $0.01$; the smallest is $0.062$. Figure~\ref{fig:shells} shows the
deviation of each shell count from its expected value, in units of its
standard deviation. The largest of the $126$ values in magnitude is $2.5$, and
the values in the five outer shells scatter about zero at every $\kappa$,
including $\kappa\le3/2$.

\subsection{Invariance under scaling and rotation}\label{sec:rotation-test}

The sampler draws $\mathbf{u}$ before applying $A$ and $Q$, so runs that start
from the same random-number seed should give, draw by draw, the same
$\mathbf{u}$ for any thermal speeds and field direction, provided $A$ and $Q$
are applied as specified. We repeated the runs at $\kappa=0.55$, $2$, and
$10$ with the same seeds for each of the six combinations of
$(\theta_\perp,\theta_\parallel)\in\{(1,1),(1,2)\}$ and field directions
$\hat{\mathbf{z}}$, $(1,1,1)/\sqrt{3}$, and the unit vector along
$(0.3,-0.5,0.8)$. Over the $7.5\times10^{8}$ draws of the five new
combinations, the normalized velocity $\mathbf{u}'$ agrees with the
velocity $\mathbf{u}$ of the corresponding draw at
$(\theta_\perp,\theta_\parallel)=(1,2)$ with the field along $\hat{\mathbf{z}}$
to $|\mathbf{u}'-\mathbf{u}|/|\mathbf{u}|\le1.0\times10^{-15}$, the level of
floating-point rounding, so the loader applies $A$ and $Q$ as specified. At
these three values of $\kappa$, the results of
Sec.~\ref{sec:cell-test} therefore hold unchanged for these thermal speeds and
field directions.

\subsection{Cartesian components and the bi-Maxwellian limit}\label{sec:marginals}

\begin{figure}[tb]
    \centering
    \includegraphics[width=\columnwidth]{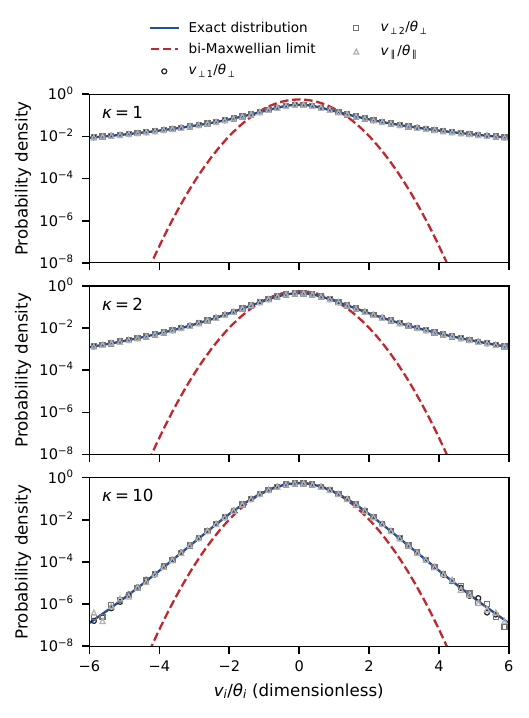}
    \caption{Distributions of the Cartesian velocity components at
    $\kappa=1$, $2$, and $10$ (top to bottom), sampled without a velocity bound
    with $(\theta_\perp,\theta_\parallel)=(1,2)$, the field along
    $\hat{\mathbf{z}}$, and $N=5\times10^{7}$ draws per panel in double
    precision. Each component is divided by its own thermal speed, so both axes
    are dimensionless. Markers are sample probability densities of
    $v_{\perp1}/\theta_\perp$ (circles), $v_{\perp2}/\theta_\perp$ (squares), and
    $v_\parallel/\theta_\parallel$ (triangles), from bins of width $0.25$
    normalized by the total number of draws, so that they can be compared with
    the curves. The solid blue curve is the exact distribution of
    Eq.~\eqref{eq:marginal}. The dashed red curve is the one-component
    distribution of the bi-Maxwellian limit, Eq.~\eqref{eq:bi-maxwell}, which
    the markers approach as $\kappa$ increases. The horizontal axis ends at
    $\pm6$ thermal speeds, which contains about $90\%$ of each component's
    distribution at $\kappa=1$ and $99.5\%$ at $\kappa=2$.}
    \label{fig:marginals}
\end{figure}

As a secondary check, we compare each Cartesian component of the returned
velocities with its exact one-dimensional distribution. Integrating Eq.~\eqref{eq:bi-kappa} over two of the three
components gives the distribution of the remaining one. For each field-aligned
component $i\in\{{\perp}1,{\perp}2,\parallel\}$, with
$\theta_{\perp1}=\theta_{\perp2}=\theta_\perp$,
\begin{equation}
p_{V_i}(v)=\frac{\Gamma(\kappa)}{\sqrt{\pi}\,\Gamma(\kappa-\tfrac12)\sqrt{\kappa}\,\theta_i}
\left(1+\frac{v^{2}}{\kappa\theta_i^{2}}\right)^{-\kappa},
\label{eq:marginal}
\end{equation}
a scaled Student-$t$ distribution with $2\kappa-1$ degrees of freedom, which is
the same for all three components in the variable $v/\theta_i$.
Figure~\ref{fig:marginals} overlays the three components on this curve, so a
wrong parallel scale would move the parallel component off it. Agreement of
the three component distributions does not determine the three-dimensional
distribution, which the tests of Sec.~\ref{sec:cell-test} examine. The
logarithmic density axis shows the agreement out to six thermal speeds and the
approach to the bi-Maxwellian limit as $\kappa$ increases. At $\kappa=1$, where
Eq.~\eqref{eq:marginal} is a Cauchy distribution and the variance does not
exist, this range contains about $90\%$ of each component's distribution; the
radius test of Sec.~\ref{sec:cell-test} covers the tail beyond it.

\subsection{Variances}

For $\kappa>3/2$, where Eq.~\eqref{eq:variances} holds, the sample variances of
the three field-aligned components give a further secondary check
(Table~\ref{table:moments}). We divide the draws into $100$ runs of
$5\times10^{5}$, compute the variance of each run, and compare the mean over
the runs with Eq.~\eqref{eq:variances}, using the standard error of that mean
as the scale. At $\kappa=5$ and $\kappa=10$ the differences
are at most $0.04\%$, within $1.6$ standard errors. At $\kappa=2$ they are
$-0.42\%$, $-0.37\%$, and $+1.71\%$. The fourth moment does not exist there, so
rare large draws can raise the variance of one run far above that of the
others and shift the mean over the runs. In the parallel component, one run
has a variance of $23.9$, against a median of $7.9$ over the runs, and this run
accounts for the $+1.71\%$. For the same reason, the standard error does not
describe the sampling error of the variance, and we do not compare these
differences with it.

\begin{table}[tb]
\caption{Sample variances of the field-aligned components against
Eq.~\eqref{eq:variances}, for $\kappa>3/2$ where the variance exists. No
velocity bound, $(\theta_\perp,\theta_\parallel)=(1,2)$, field along
$\hat{\mathbf{z}}$, double precision. The sample column gives the mean of the
variances of $100$ runs of $5\times10^{5}$ draws $\pm$ its standard error. At
$\kappa=2$ the fourth moment does not exist, so the standard error does not
describe the sampling error of the variance and is not given.}
\label{table:moments}
\centering
\begin{ruledtabular}
\begin{tabular}{llccr}
$\kappa$ & & Expected & Sample & Diff.\ (\%) \\
\hline
$2$ & $V_{\perp1}$ & 2.0000 & 1.9916 & -0.42 \\
$2$ & $V_{\perp2}$ & 2.0000 & 1.9925 & -0.37 \\
$2$ & $V_\parallel$ & 8.0000 & 8.1369 & +1.71 \\
\hline
$5$ & $V_{\perp1}$ & 0.7143 & 0.7146 $\pm$ 0.0002 & +0.04 \\
$5$ & $V_{\perp2}$ & 0.7143 & 0.7143 $\pm$ 0.0002 & 0.00 \\
$5$ & $V_\parallel$ & 2.8571 & 2.8563 $\pm$ 0.0008 & -0.03 \\
\hline
$10$ & $V_{\perp1}$ & 0.5882 & 0.5882 $\pm$ 0.0001 & -0.01 \\
$10$ & $V_{\perp2}$ & 0.5882 & 0.5882 $\pm$ 0.0001 & 0.00 \\
$10$ & $V_\parallel$ & 2.3529 & 2.3531 $\pm$ 0.0004 & +0.01 \\
\end{tabular}
\end{ruledtabular}
\end{table}

\section{Scope and limitations}\label{sec:discussion}

\emph{Scope of the validation.} The distribution tests used
$\theta_\parallel/\theta_\perp=2$ with the field along a coordinate axis;
other thermal-speed ratios and field directions were checked through the
invariance of the normalized velocity at three values of $\kappa$
(Sec.~\ref{sec:rotation-test}). In single precision, and for $\kappa<0.51$,
Sec.~\ref{sec:precision} measured the lost draws but not the distribution of
the returned ones. For other parameters the released test program can be
applied to the actual load.

\emph{Finite precision.} The unavoidable loss follows from the target
distribution and the floating-point format alone. A loader can compare its
measured loss with this value; an excess indicates a loss in the calculation.

\emph{The cap.} The tail was characterized by a single statistic, the
$99.9$th-percentile speed; a quantity weighted further into the tail may
change more under the same cap. The box also breaks the rotational symmetry
of the distribution about the field. Its orientation in the plane
perpendicular to the field is set by the choice of
$\hat{\mathbf{e}}_{\perp1}$ (Sec.~\ref{sec:transform}), so where the field
direction varies, this orientation can change abruptly between neighboring
regions. The rejected fraction and the speed distribution do not depend on
it. A cap makes every moment finite, but the values depend on $\lambda$, a
numerical parameter. When finite moments are wanted for physical reasons,
the regularized Kappa distribution (Sec.~\ref{sec:three-laws}) makes the
cutoff a parameter of the model, and samplers for it
exist~\cite{zenitani2026nonmaxwellian}.

\section{Summary}\label{sec:summary}

We tested the complete output of an anisotropic bi-Kappa loader built on the
Gamma-ratio construction~\cite{abdul2015bernstein,zenitani2022relativistic,
zenitani2023losscone,zenitani2026nonmaxwellian}.

Near $\kappa=1/2$, the direct calculation of the radius loses large-radius
draws when the floating-point product that forms its denominator rounds to
zero. Computing the radius on a logarithmic scale avoids this loss. The loss
that remains depends only on the target distribution and the floating-point
format, so it can be computed before a run, and a loader built on another
Gamma generator can compare its measured loss with it.

The rejected fraction of a component-wise bound decays only as
$\lambda^{-(2\kappa-1)}$, and a small rejected fraction can accompany a large
change in a tail percentile. The region and width of a bound are therefore
part of the loaded distribution and should be reported with $\kappa$,
$\theta_\perp$, and $\theta_\parallel$.

Tests of the normalized radius, the direction, and their independence need no
finite moments, so they apply for $1/2<\kappa\le3/2$, where the second moments
do not exist, and they reach far into the tail. The loaded velocities were
consistent with the target distribution at every tested $\kappa$. The released
program applies these tests to velocities from any loader.

\section*{Code availability}

The loader and the test program are released as
\texttt{bi-kappa-distribution-sampling} under the MIT License at
\href{https://github.com/KehengZhu/bi-kappa-distribution-sampling}{\nolinkurl{github.com/KehengZhu/bi-kappa-distribution-sampling}}
and deposited on Zenodo at
\href{https://doi.org/10.5281/zenodo.20617011}{\nolinkurl{doi.org/10.5281/zenodo.20617011}}~\cite{bikappa_code};
the DOI points to the latest deposited version, and the version evaluated here
is 3.0.0. The package contains C++ samplers for the bi-Kappa and bi-Maxwellian
distributions, with or without a component-wise velocity bound, and a Python
program that applies the tests of
Sec.~\ref{sec:validation} to velocities from any loader and, given the number
of attempts, reports the rejected fraction. Without a bound, the samplers
return a draw with a non-finite component rather than discarding it, so that
the number of lost draws is known. The samplers are used in the HYPERS
hybrid particle simulation
code~\cite{omelchenko2012hypers,omelchenko2023agu,omelchenko2023emaps}.

\bibliography{refs}

\end{document}